\documentclass{article}
\usepackage{amssymb}

\input{tcilatex}

\begin{document}

\title{Unusual Features of Varying Speed of Light Cosmologies}
\author{John D. Barrow \\
%EndAName
DAMTP, \\
Centre for Mathematical Sciences,\\
Cambridge University\\
Cambridge CB3 0WA\\
UK}
\maketitle

\begin{abstract}
We contrast features of simple varying speed of light (VSL) cosmologies with
inflationary universe models. We present new features of VSL cosmologies and
show that they face problems explaining the cosmological isotropy problem.
We also find that if $c$ falls fast enough to solve the flatness and horizon
problems then the quantum wavelengths of massive particle states and the
radii of primordial black holes can grow to exceed the scale of the particle
horizon. This may provide VSL cosmologies with a self-reproduction property.
The constraint of entropy increase is also discussed. The new problems
described in the this letter provide a set of bench tests for more
sophisticated VSL theories to pass.
\end{abstract}

\section{\protect\bigskip Introduction}

We have explored a naive class of models in which the speed of light varies
in time \cite{am}, \cite{jbvsl}. These can be reformulated more generally
and recast as theories in which other dimensional 'constants' carry the
operationally meaningful space or time variation of a dimensionless constant
-- in this case the fine structure 'constant', $e^{2}/\hbar c,$ \cite%
{barmag98}. Theories in which this variation is carried explicitly by the
electron change have been extensively investigated recently \cite%
{bsbm,bsm1,bsm2,bsm3,bsm4,bm,bmaj}. Such theories are a 'small' perturbation
of standard physics in the sense that even though the speed of light falls
there still exists a maximum signal propagation velocity that is achieved by
gravitational waves. The motivation for a careful consideration of these
cosmological models was their possible viability as alternatives to
inflation as explanations for a number of unusual properties of the universe
and the consistency of quasar absorption spectra with a variation in the
value of the fine structure 'constant at $z\sim 1-3,$ \cite{murphy, webb,
webb2,webb3}. In particular, it was shown by Moffat \cite{moffatal, mof},
Albrecht and Magueijo \cite{am} and by Barrow \cite{jbvsl} that during the
very early universe a finite period of time during which the speed of light
falls at an appropriate rate can lead to a solution of the flatness,
horizon, and monopole problems. However, unlike inflation it can also
provide a solution for the cosmological constant problem. Barrow and
Magueijo also showed that for some ranges of variation in the speed of light
these theories can also naturally create long-lived universe in which the
dynamics are almost flat or in which the cosmological constant is almost
zero. These problems of explaining a universe in which the present value of
the matter density parameter, $\Omega _{m}$, or the cosmological constant
energy density parameter, $\Omega _{\Lambda }$, are $O(0.1-1)$ we called the
quasi-flatness and quasi-lambda problems. Inflation does not seem to offer a
natural explanation of a universe which is almost flat or has a dynamically
significant cosmological constant today, as current observations imply.

In this paper, we want to discuss some other cosmological problems in the
light of varying speed of light\ (VSL) theories of the simple sort discussed
by Albrecht, Magueijo, and Barrow (AMB). Although these problems are
articulated in the context of this simple AMB theory we believe that are
more general challenges to any theory which manifests itself in a particular
coordinate system as a VSL theory. In particular, we shall show that in
general AMB theories cannot solve the isotropy problem and are unable to
generate a spectrum of almost constant curvature fluctuations in the
standard way. We shall also show that they have a number of very unusual
consequences for the quantum states of massive particles and for primordial
black holes after they enter the particle horizon in the early universe. It
is not clear whether these features are a reductio ad absurdum for these VSL
theories or whether they provide an exotic counterpart to the
self-reproducing property of inflationary universe models that provides the
basis for eternal inflationary universes.

\section{Naive VSL Theories}

\subsection{Solving the flatness, horizon and lambda problems}

The simple VSL model introduced by Albrecht and Magueijo and solved by
Barrow is based upon the simplest possible premises. We assume that the
Friedmann cosmology in the presence of time-dependent speed of light $c(t)$
is described by the Friedmann and Raychaudhuri equations (although the
original formulation allowed the Newtonian gravitation 'constant' $G$ to
vary also, we shall ignore this possibility as it makes no contribution to
the essential conclusions):

\begin{equation}
\frac{\ddot{a}}{a}=-\frac{4\pi G}{3}(\rho +\frac{3p}{c^{2}(t)})+\frac{%
\Lambda c^{2}(t)}{3}  \label{ray}
\end{equation}

\begin{equation}
\frac{\dot{a}^{2}}{a^{2}}=\frac{8\pi G\rho }{3}-\frac{kc^{2}(t)}{a^{2}}+%
\frac{\Lambda c^{2}(t)}{3}  \label{frw}
\end{equation}
where $a(t)$ is the expansion scale factor of the Friedmann metric, $p$ is
the fluid pressure, $\rho $ is the fluid density, $k$ is the curvature
parameter, $\Lambda $ is the cosmological constant, and all derivatives are
with respect to $t$, the comoving proper time. From these equations we can
derive the modified matter conservation which incorporates the effects of $%
\dot{c}\neq 0:$

\begin{equation}
\dot{\rho}+\frac{3\dot{a}}{a}(\rho +\frac{p}{c^{2}(t)})=\frac{3kc^{2}\dot{c}%
}{4\pi Ga^{2}}  \label{cons}
\end{equation}

Of course, in general relativity, where $\dot{c}=0,$ the right-hand side of (%
\ref{cons}) is zero. Notice that it also vanishes in a VSL theory when $k=0.$
For concreteness, now consider the universe to contain a perfect fluid with
equation of state

\begin{equation}
p=(\gamma -1)\rho c^{2}  \label{p}
\end{equation}
where $\gamma $ is constant and we shall assume that the speed of light
varies as some power of the expansion scale factor:

\begin{equation}
c(t)=c_{0}a^{n}  \label{c}
\end{equation}
where $c_{0}>0$ and $n$ are constants. These assumptions are sufficient to
solve the equations completely. Integrating (\ref{cons}) with $\Lambda =0$,
we find

\begin{equation}
\rho =\frac{M}{a^{3\gamma }}+\frac{3kc_{0}^{2}na^{2n-2}}{4\pi G(2n-2+3\gamma
)}  \label{rho}
\end{equation}
Now by inspection of (\ref{frw}) we see that at large $a$ the curvature term
falls off faster than the $Ma^{-3\gamma }$term whenever

\begin{equation}
n<\ \frac{1}{2}(2-3\gamma )  \label{n}
\end{equation}%
Thus the flatness problem can be solved in a radiation-dominated early
universe by an interval of VSL evolution if

\[
n<-1 
\]

It is easy to show that, as in the case of inflation, this is the same
condition required to solve the monopole and horizon problems. However, the
nature of the solution differs from that provided by a period of inflation.
Inflation requires the dynamics to be dominated for a finite period by an
unusual fluid ($\gamma <-\frac{2}{3}$), which is therefore gravitationally
repulsive, so that the expansion will accelerate ($\ddot{a}>0$) and the $%
Ma^{-3\gamma }$term will grow to dominate the $kc^{2}a^{-2}$ term in (\ref%
{frw}) as $a(t)$ grows large. By contrast, in the VSL model, no special
fluid is required and in a radiation universe the flatness problem is solved
because the $kc^{2}a^{-2}$ term falls off faster than the $Ma^{-3\gamma }$%
term because of the fall in $c(t)$ as $a(t)$ increases.

This behaviour also permits a solution of the classical cosmological
constant problem in just the same way if

\[
n<-\frac{3\gamma }{2} 
\]
When $\Lambda >0$ we can solve the system of equations to obtain the density:

\[
\rho =\frac{M}{a^{3\gamma }}+\frac{3kc_{0}^{2}na^{2n-2}}{4\pi G(2n-2+3\gamma
)}-\frac{\Lambda c_{0}^{2}na^{2n}}{4\pi G(2n+3\gamma )} 
\]

In this case the $\Lambda c^{2}$ term falls off faster with increasing time
than the $Ma^{-3\gamma }$term and the dynamics naturally approach those of
the $k=0=\Lambda $ Friedmann model at late times. A period of
radiation-dominated evolution with

\[
n<-2 
\]
can therefore solve an initial classical lambda problem.

Inflationary universe models do not offer a solution of this problem. It is
interesting that the flatness and cosmological constant problems can be
solved by these VSL theories within any fine tuning of $n$. But note also
that even in the VSL models there is no solution of the non-classical $%
\Lambda $ problem because a sequence of phase transitions are able to
reinstate the value of the cosmological constant at successive cosmological
epochs even if its value is set equal to zero at some arbitrarily early time.

The simple mathematical model for VSL considered here is oversimplified in
many respects but it can easily be put on a former foundation. If we define
a scalar field $\psi =\log (c/c_{0})$ then a simple action to describe its
coupling to gravity is the choice

\[
S=\int d^{4}x\sqrt{-g}\left( e^{a\psi }(R-\kappa (\psi )\nabla _{\mu }\psi
\nabla ^{\mu }\psi )+\frac{16\pi G}{c_{0}^{4}}e^{b\psi }L_{m}\right) 
\]%
parametrized by the constants $a,b,$ and $\kappa .$ Similar properties of
the resulting cosmological solutions are found to those displayed by the
simple theory above. Many of the potential problems raised in the paper need
to be readdressed in the context of a specific theory, defined by a given
lagrangian. At present there is no unique VSL theory. However, the naive VSL
theory described above plays an important role in generating hypotheses to
be tested in more sophisticated theories.

\subsection{\protect\bigskip The Quasi-flatness and Quasi-lambda Problems}

However, although string theories seem to predict that the cosmological
constant should be zero, there is solid observational evidence that it is \
small and positive and dominating the expansion dynamics of the universe
today ($\Omega _{m}\sim 0.3,\Omega _{\Lambda }\sim 0.7$). These observations
lead us to ask whether there can be 'natural' explanations for universes
which are 'almost' flat or have 'almost' zero cosmological constant. Barrow
and Magueijo \cite{bm} have shown that these simple VSL models can provide
solutions to both problems if the range of $n$ is narrowed. Thus, when the
speed of light falls as (\ref{c}) with

\begin{equation}
\ 0>n>\frac{1}{2}(2-3\gamma )  \label{cond}
\end{equation}%
we find that at late times we have a quasi-flat universe and

\[
\Omega \rightarrow \frac{-2n}{3\gamma -2} 
\]
Similarly, when

\[
0>n>-\frac{3\gamma }{2} 
\]
we have a significant cosmological constant contribution to the expansion at
late times with

\[
\frac{\Omega _{m}}{\Omega _{\Lambda }}\rightarrow -\frac{3\gamma \Omega _{m}%
}{2(\Omega _{\Lambda }+\Omega _{m})}. 
\]

\section{Other Consequences of VSL Theories}

\subsection{The Isotropy Problem}

The isotropy problem is solved by inflationary universe models with $0\leq
\gamma <2/3$ because the dominant anisotropy modes in expanding anisotropic
universes fall off no slower than $\sigma ^{2}\varpropto a^{-2}$ and
equation (\ref{frw}) is modified in the presence of anisotropies by the
addition of a\ $\sigma ^{2}$ term to its right-hand side. Thus we see that
at late times in an inflationary expansion the $Ma^{-3\gamma }$term falls
off more slowly than $\sigma ^{2}\varpropto a^{-2}$ and dominates the
dynamics, driving the expansion away from isotropy. Note however that in
generic ever-expanding universes the source of this dominant $\sigma
^{2}\varpropto a^{-2}$ or $t^{-2}$ late-time anisotropy is the anisotropy in
the 3-curvature of space. The simpler, much-studied anisotropy
characteristic of Bianchi I or Kasner universes, arises from simple
expansion rate anisotropy with isotropic spatial 3-curvature is sub-dominant
at late times, with $\sigma ^{2}\varpropto a^{-6}.$ This behaviour is of
measure zero in the space of all anisotropic cosmological models. A solution
of the anisotropy problem must explain why the $\sigma ^{2}\varpropto a^{-2}$
anisotropy mode has not come to dominate the expansion. This is possible for
inflationary expansion dominated by a $\rho +3p<0$ effective stress since
this falls off more slowly than $\sigma ^{2}$ as $t\rightarrow \infty $ and
the influence of the anisotropy can be made arbitrarily small for a
sufficiently long period of accelerated expansion.

In the situation, with constant $c$, where inflation does not occur and
isotropisation is not generic. The anisotropising mode is well understood in
the context of homogeneous and anisotropic cosmologies. It is a Bianchi type 
$VII_{h}$ plane gravitational wave described by an exact solution found by
Lukash \cite{luk}. It is interesting to note that they can be excluded if an
open universe has a finite topology \cite{kod}$\emph{.}$In the case with the
natural $R^{3}$ topology one can show that a particular family of known
exact vacuum solutions of type $VII_{h\text{ \ \ }}$are stable as $%
t\rightarrow \infty $ \cite{jbaniso, qj, grg}. In particular, the isotropic
open Friedmann universe is stable (but not asymptotically stable) with
respect to these anisotropic curvature modes: that is, as $t\rightarrow
\infty $, the ratio of the shear $\sigma $ to mean Hubble expansion rate, $H=%
\dot{a}/a,$ approaches a non-zero constant value. Inflation can make this
constant arbitrarily close to zero.\emph{\ }

In the simplest anisotropic universes with comoving perfect-fluid matter and
isotropic 3-curvature and VSL we can repeat the approach taken in the case
of isotropic universes, using the Raychaudhuri acceleration equation, its
first integral, and the shear propagation equation -- with $c(t)$ assumed --
to derive a generalised matter conservation equation. Thus

\[
\frac{d}{dt}\left( \frac{3H}{c}\right) =-3\frac{H^{2}}{c^{2}}-2\frac{\sigma
^{2}}{c^{2}}-\frac{4\pi G}{c^{4}}(\rho +3p)+\Lambda 
\]

\begin{equation}
^{3}R=\frac{16\pi G\rho }{c^{4}}-\frac{6H^{2}\ }{c^{2}}+\frac{2\sigma ^{2}}{%
c^{2}}+2\Lambda  \label{fr}
\end{equation}

\[
\ \frac{d}{\ dt}\left( \frac{\sigma }{c}\right) \ +3H\frac{\sigma }{c}=0 
\]

Hence we see that $\sigma =\Sigma ca^{-3}$ and if we were to write $%
^{3}R=2ka^{-2}$ then

\[
\dot{\rho}+3H(\rho +\frac{p}{c^{2}})+\frac{c\dot{c}}{4\pi G}\left( \frac{\
2\ \Sigma ^{2}}{a^{6}}-\frac{3k\ }{\ a^{2}}\right) \ =0 
\]%
Using eqns. (\ref{p}) and (\ref{c}) we can solve for $\rho $ and we find
that the flatness problem is solved as usual when (\ref{n}) holds. We can
see immediately that the shear effects are negligible with respect to the
curvature and

\bigskip

\[
\rho =\frac{\Gamma }{a^{3\gamma }}-\frac{2\Sigma ^{2}nc_{0}a^{2n+3\gamma -6}%
}{(2n+3\gamma -6)}+\frac{6knc_{0}^{2}a^{2n+3\gamma -2}}{(2n+3\gamma -2)}. 
\]%
Hence, the shear term is always insignificant and the condition for solution
of the flatness problem is the same as in an isotropic universe, (\ref{cond}%
).

In the general case we have to consider the contribution made the
anisotropic 3-curvature terms and anisotropic fluid pressure and not just an
anisotropic Hubble flow ($\sigma \neq 0$), as considered above. For
simplicity, we neglect the anisotropic pressure effects but consider the
role of anisotropic 3-curvature. The shear propagation equation becomes \cite%
{ellis, vanel}, :

\[
\frac{h_{\rho }^{\mu }h_{\sigma }^{\nu }}{c}\frac{d}{dt}\left( \frac{\sigma
^{\rho \sigma }}{c}\right) =-\frac{2H\sigma ^{\mu \nu }}{c^{2}}-\frac{\sigma
_{\rho }^{\mu }\sigma ^{\nu \rho }}{c^{2}}-E^{\mu \nu }+\frac{2\sigma
^{2}h^{\mu \nu }}{3c^{2}} 
\]

\[
E_{\mu \nu }^{{}}=\ S_{\mu \nu }+\frac{\ H\sigma _{_{\mu \nu }}^{\ \ }}{c^{2}%
}-\frac{\sigma _{\mu \rho }\sigma _{\nu }^{\rho }}{c^{2}}+\frac{2\sigma
^{2}h_{\mu \nu }}{3c^{2}} 
\]%
where $S_{\mu \nu }$ is the anisotropic part of the spatial 3-curvature and $%
h_{\mu \nu \text{ \ }}$is the projection tensor. Generic evolution at late
times for ever-expanding anisotropic universes close to isotropy with $%
\Lambda =0$ will have the scaling form $S_{\mu \nu }\sim E_{_{^{\mu \nu
}}}\sim t^{-2},Hc^{-1}\sim t^{-1},H\sigma c^{-2}\sim t^{-2},\sigma
c^{-1}\sim t^{-1}.$ Hence we $\sigma \propto H$ as in the case with constant 
$c$. However, we see that $\sigma ^{2}c^{-2}\sim t^{-2}$ in the generalised
Friedmann equation (\ref{fr}) whereas the isotropic matter terms go as $\rho
c^{-4}\propto a^{-3\gamma -4n}.$Thus the shear terms will dominate at late
times close the Friedmann expansion, $a\propto t^{2/3\gamma }$ whenever $n<0$
and a falling speed of light does not solve the anisotropy problem. There
are several ways in which the VSL effects on anisotropy can be understood
more physically. The naive VSL theories preserve the metric structure of
spacetime so that gravitational-wave propagation still occurs at a maximum
possible propagation speed. But light propagates with a variable speed that
is less than or equal to the gravitational-wave propagation speed. Thus one
can see that anisotropies that are carried by long-wavelength gravitational
waves can avoid being made innocuous by a fall in the speed of light. The
generic anisotropies at late time in ever-expanding Bianchi types that
contain isotropic Friedmann universes are of this type.

\subsection{The Inhomogeneity Problem}

The spectrum of primordial fluctuations is the most interesting prediction
that any VSL model can make because the observational evidence from the
microwave background temperature fluctuations is potentially the most
decisive of observational tests. Inflationary theories have been able to
provide a natural explanation for our observations of an almost constant
curvature spectrum of inhomogeneous fluctuations in the universe. Other
competing predictions have been made by colliding brane models \cite{brane}.

In inflationary universes the approximately constant curvature spectrum of
inhomogeneities arises from the near de Sitter behaviour of the expansion
dynamics during an inflationary epoch where $\gamma \approx 0.$ In a finite
time interval of $c$ variation in a VSL theory that solves the flatness or
lambda problems the expansion dynamics will be dominated by the usual $%
Ma^{-3\gamma }$term. If the expansion is radiation dominated then no special
inhomogeneity spectrum of the constant curvature form will be created by the
VSL evolution. One way of imprinting a characteristic spectrum could be via
the sudden phase transition model of VSL favoured by Moffat and Albrecht and
Magueijo. Here, it would need to be shown that a constant curvature spectrum
results. Again, this is a major challenge because the phase transition
models of inflation achieve a constant curvature spectrum of fluctuations by
virtue of their proximity to a de Sitter state in the vacuum state where the
scalar field stops rolling. Recently, one attempt along these lines for the
creation of a constant curvature spectrum has been made by Moffat \cite{cc},
and Magueijo and Pogosian \cite{mp} have investigated the possibility that
thermal fluctuations might give rise to an interesting primordial
inhomogeneity spectrum as a result of modified dispersion relations or
cosmological bounce at high energy.

\subsection{The Massive Particle Problem}

VSL theories have two further strange properties that appear to have
dramatic consequences. Suppose that the universe starts to experience a
period of VSL evolution of the form (\ref{c}). Let us consider the fate of a
particle state of mass $M$ that exists on a scale smaller than the particle
horizon $R_{hor}\sim ct$. Now if it has a de Broglie wavelength

\[
\lambda =\frac{\hbar }{Mc} 
\]
then if $c$ falls as $c\varpropto a^{n}\varpropto t^{n/2}$ in a
radiation-dominated universe we have for the ratio of $\lambda $ to the
proper size of the particle horizon

\[
R_{hor}=a(t)\int \frac{cdt}{a(t)} 
\]
is given by

\[
\frac{\lambda }{R_{hor}}\approx \frac{\hbar }{Mc}\frac{(n+1)}{%
2c_{0}t^{(n+2)/2}}=\frac{(n+1)\hbar }{2Mc_{0}^{2}t^{(2n+2)/2}}\varpropto
t^{-n-1}\rightarrow \infty 
\]%
as $t\rightarrow \infty $ if $n<-1.$ So if we take $n<-1$ as required to
solve the flatness problem, massive particles evolve to become acausal
separate universes! Note that in particular theories $\hbar $ may vary as
well. For example in the naive VSL theory of refs \cite{am}, \cite{jbvsl} $%
\hbar \varpropto c\varpropto a^{n}$ and so $\lambda /R_{hor}\varpropto t^{-%
\frac{n}{2}-1}$ and a massive particle problem can arise when $n<2,$ as
required to solve the classical cosmological constant problem.

How should be interpret this. Is is a reductio ad absurdum of the VSL or
should we take a lesson from new inflation and regard our observable
universe as the interior of the single particle state. This might provide a
natural explanation for some of its coordinated properties and provide a
reason for a special irregularity spectrum to be formed. Note that this
behaviour of massive particle states ceases when the VSL evolution ends and
will not be going on in the universe today. Since the early universe may
contain a population of massive states inside the horizon scale when VSL
evolution begins, the ensuing evolution is not dissimilar to the
self-reproducing inflationary universe. There small regions create eternal
inflation due to the quantum evolution dominating the classical slow roll
under fairly general conditions. The rapid expansion of many VSL regions
would produce collisions that might lad to unacceptable levels of
inhomogeneity when they subsequently re-enter the horizon if the amount of
expansion was small. But if it was large then we could find ourselves
inhabiting a single VSL pre-expanded domain. This scenario may repay further
detailed analysis. There are many complexities that have been ignored here,
in particular relating to the rapid growth of dimensionless couplings like $%
e^{2}/hc,$ $g^{2}/hc,$ $Gm^{2}hc$ as $c$ falls, rendering all interactions
strong. It may be that the avoidance of this problem is a constraint that
needs to be placed on sophisticated VSL theories or it could be exploited as
a new mechanism for making small local regions become large.

\subsection{The Primordial Black Hole Problem}

A similar fate awaits any small primordial black hole of mass $M$ that forms
on sub-horizon scales. Let us ignore Hawking evaporation and consider the
ratio of the Schwarzschild radius of the black hole, $R_{bh\text{ \ }}$to
the horizon scale:

\[
\frac{R_{bh}}{R_{hor}}=\frac{2GM}{c^{2}}\frac{(n+1)}{2c_{0}t^{(n+2)/2}}%
\varpropto t^{-1-\frac{3n}{2}} 
\]%
So if $n<-2/3$ the black hole horizon grows faster than the particle horizon
and the black hole becomes an acausal separate universe. Much of the
discussion regarding the expansion of massive particle states applies to
this situation also. Here there is an added interpretational uncertainty in
that it is not clear what happens to the black hole when $c$ changes.
Changing $c$ may be sufficient to stop a black hole forming. But if a
Schwarzschild black hole formed when $c$ did not change then it might be
that it had to remain constant on the horizon even while $c$ changed in the
background.

\subsection{\protect\bigskip An Entropy Problem}

If we believe that we can apply the Bekenstein-Hawking entropy formula to
the particle horizon of an expanding universe then the entropy inside the
VSL horizon is

\[
S\varpropto \frac{R_{hor}^{2}}{R_{pl}^{2}}\ 
\]%
and this entropy evolves as $t^{\frac{5n+4}{2}}$in VSL theories and hence
increases only if $n>-4/5$ in the radiation era when $\hbar $ is constant.
If $\hslash \propto c\propto a^{n}$, so $R_{pl}\propto \ a^{-n},$ then the
entropy increases only if $n>-1$ $.$ In general we might consider a suite of
theories in which $\hbar c\propto a^{\lambda }$ so $R_{pl}\propto
a^{(\lambda -4n)/2}$ so $S\propto a^{6n-\lambda }t^{2}$ and entropy
increases during the radiation era when $3n+2-\lambda /2>0.$ Thus the
solution of the flatness an lambda problems can require this entropy measure
to decrease with increasing time. The Bekenstein-Hawking entropy technically
applies only to event horizons but many analogous measures of 'gravitational
entropy' have been proposed and it is possible that the argument framed here
will have application to any better-motivated masure of gravitational
entropy. This worry also besets arguments like those of Davies et al \cite%
{dav} which use black hole thermodynamics to assess whether the variation of
certain constants are in accord with the second law of thermodynamics. The
difficulty is that the required black holes and their thermodynamic will
only exist as particular solutions of a theory with varying constants --
particular solutions in which those varying constants take constant values
-- and so the argument cannot be carried through. A specific case arises in
Brans-Dicke theory. The Schwarzschild solution is a particular solution of
Brans-Dicke theory and its Bekenstein-Hawking entropy is $S\varpropto GM^{2}$%
. One might be tempted therefore to think that any solution of Brans-Dicke
gravity in which $G$ decreases with time would therefore violate a second
law of black hole thermodynamics. However, the Schwarzschild black hole is
only a solution of Brans-Dicke theory when $G$ is constant. If $G$ is
allowed to vary (as in the thought $\ $experiment designed to violate the
second law) the static spherically symmetric solution of the Brans-Dicke
equations is not a black hole.

\section{Discussion}

We have considered further properties of naive VSL cosmologies, in addition
to those related to the flatness and horizon problems. There are problems
explaining the isotropy of the universe in general and unusual consequences
of varying $c$ on particle de Broglie wavelengths and black holes. It is not
clear whether the massive particle and black hole problems are fatal to the
conception of VSL theories. At first sight they appear to create an
unsatisfactory state of affairs. However, it may be that they are in effect
counterparts of the self-reproduction property of inflationary universes
that gives rise to the quasi-stationary eternal inflationary universe
scenario. The rapid enlargement of single particle states to super-horizon
scales simply provides a way of producing large scale regions that possess
the coherent properties possessed by single particle states when they were
on sub-horizon scales and to generate fluctuations on super-horizon scales.
However, although states grow faster than the horizon they do not
necessarily grow fast enough to create regions which could encompass our
entire visible universe. Whether or not these features provide a reductio ad
absurdum for these cosmologies remains to be explored in greater detail. Any
more completely specified VSL theory will need to be judged both by its
ability to solve the standard problems that inflation can resolve and by its
ability to circumvent the difficulties described in this letter.

Acknowledgements. I would like to thank R. Maartens, J. Magueijo and J.
Moffat for discussions.

\end{document}